# Derivation of Spin Torques of the Magnetic System in Broken Inversion Symmetry

H. Y. Kwon, S. P. Kang, N. J. Kim, C. K. Lee and C. Won

***Department of Physics, Kyung Hee University, Seoul 02447, Korea***

We propose a new approach to derive spin torque in systems of broken inversion symmetry. It uses the concepts of asymmetric and directional spin-spin interactions to obtain their effective fields. We applied the effective fields into the Landau-Lifshitz equation and obtained spin torques. The model offers a new and general approach for spin dynamics, one that effectively merges the Dzyaloshinskii-Moriya interaction, spin transfer torques, and spin-orbit torque into the spin dynamics equation. We discussed how our model is imposed on the spin dynamics and compared our approach with the traditional discussions on spin dynamics.

# I. INTRODUCTION

In recent years, the magnetic properties and dynamics in the system of broken inversion symmetry have become one of the most intensely investigated topics in magnetism studies. When the magnetic moments are located in the environment of broken inversion symmetry or when the directional preference is applied to the system, the energy and the torque on the magnetic moments should include various additional contribution such as the Dzyaloshinskii-Moriya interaction (DMI), spin transfer torque (STT), spin-orbit torque (SOT) from Rashba effect, spin hall (SH) effect, and etc.[1-12]. The broken inversion symmetry may arise from the spin current as well as the crystalline structure. In the case of DMI, the inversion symmetry is broken by the structure such as non-centrosymmetric crystalline structure, surface, or interface with heavy metal. In the case of STT, the inversion symmetry is broken by the direction of spin current. Among them, the DMI can be effectively applied in spin dynamics using its effective field, similarly as other interactions such as exchange interaction and anisotropy are. However, the other contributions STT and SOT have been discussed with explicit torque terms not derived from effective fields.

In this paper, we will discuss a new approach to model the dynamics of spin-spin interactions in broken inversion symmetry environment by introducing asymmetric and directional interaction into the spin dynamics equation. We obtained the effective fields causing spin torques directly from the asymmetric and directional interaction and found the spin torques have the same form of spin transfer torque and spin orbit torque. We discuss how the above consideration affects the spin dynamics of the system and compare how our model fits into other traditional approaches.

# II. THEORY

The spin-spin interaction energy in magnetic system is written as $-\sum_{<i,j>}\vec{S}_i\cdot\overleftrightarrow{J}\cdot\vec{S}_j$, where $\overleftrightarrow{J}$ is a general exchange interaction tensor, $\vec{S}_i$ and $\vec{S}_j$ are normalized spin vectors. When inversion symmetry exists in the magnetic system, the energy term can be simplified as $-J\sum_{<i,j>}\vec{S}_i\cdot\vec{S}_j$, which is usually called the exchange interaction energy, $\mathcal{H}_{\mathrm{ex}}$, where the $J$ is a parameter related with exchange interaction strength. For the exchange interaction case, the effective field, $\vec{h}_{i,J}(=-\frac{1}{\mu_0 m}\frac{\partial\mathcal{H}_{\mathrm{ex}}}{\partial\vec{S}_i})$, which is commonly used scheme to study magnetic system, can be written by $\frac{J}{\mu_0 m}\sum_j^{nn}\vec{S}_j$, where $m$ is the constant magnitude of the magnetic moment and $\mu_0$ is the magnetic permeability of vacuum. If a spin is in an environment with broken inversion symmetry, the DMI is involved, adding the energy term $\mathcal{H}_{\mathrm{DM}}=-\sum_{<i,j>}\vec{D}_{ij}\cdot(\vec{S}_i\times\vec{S}_j)$[13,14]. DMI corresponds to the off-diagonal term in the general exchange interaction tensor $\overleftrightarrow{J}$, and its effective field can be written by $\vec{h}_{i,\vec{D}}=\frac{1}{\mu_0 m}\sum_j^{nn}(\vec{S}_j\times\vec{D}_{ij})$.

Additionally, we propose that one should consider the asymmetric or directional preference of the spin-spin interaction in the case of broken inversion symmetry; a spin may interact more on one side than the other side, as presented in Fig. 1. Actually, the exchange interaction is related to $\vec{S}_i\cdot\vec{S}_j=S_{i,z}S_{j,z}+\frac{1}{2}(S_i^+S_j^-+S_i^-S_j^+)$, and the later terms involving spin ladder operators are related to spin transfer from *j* to *i*, or from *i* to *j*. The asymmetry in these terms will cause a directional spin flow, and it is proportional to the current density and the strength of spin polarization of the current from the broken inversion symmetry. For example, assuming there is a spin current flowing in one direction, the interaction strengths will have a specific directional preference, and the interactions from left to right and from right to left may no longer be identical. Another example would be spins on the edge of the system, which do not have neighboring spins in the same direction. The magnetoelastic effects, which are related to the lattice vibration, are another example of the directional difference of an exchange interaction[15]. As represented in Fig. 1(a), we considered that the exchange constants $J$ differ by $\delta J$, which can be positive or negative according to the broken symmetry characteristics.

The effective field caused by the exchange interaction with *x*-directional broken symmetry is given by

$$
\begin{aligned}
\vec{h}_{i,\mathrm{ex}} &= \frac{1}{\mu_0 m}\left(\left(J+\frac{\delta J}{2}\right)\vec{S}_{i+x}+\left(J-\frac{\delta J}{2}\right)\vec{S}_{i-x}\right)\\
&\cong \frac{1}{\mu_0 m}\left(\left(J+\frac{\delta J}{2}\right)\left(\vec{S}_i+\frac{\partial \vec{S}_i}{\partial x}a+\frac{1}{2}\frac{\partial^2 \vec{S}_i}{\partial x^2}a^2\right)+\left(J-\frac{\delta J}{2}\right)\left(\vec{S}_i-\frac{\partial \vec{S}_i}{\partial x}a+\frac{1}{2}\frac{\partial^2 \vec{S}_i}{\partial x^2}a^2\right)\right) \qquad (1)\\
&= \vec{h}_{i,J}+\frac{a\,\delta J}{\mu_0 m}\frac{\partial \vec{S}_i}{\partial x},
\end{aligned}
$$

where $a$ is the unit length of the system, and the additional effective field term $\delta\vec{h}_{i,\delta J}\left(=\frac{a\,\delta J}{\mu_0 m}\frac{\partial \vec{S}_i}{\partial x}\right)$ is the leading additional effective field of the asymmetric interaction. Since we used normalized spin vector, the magnitude of $\vec{S}_i$ is constant, and $\frac{\partial \vec{S}_i}{\partial x}$ is perpendicular to the direction of $\vec{S}_i$. Therefore, the direction of the additional effective field is perpendicular to the spin direction, and the spin cannot be at rest unless the strength of $\delta\vec{h}_{i,\delta J}$ is zero. Thus, this effective field results in a dynamic feature of spins, such as the domain wall (DW) motion or spin wave.

Naturally, we can consider the asymmetric DMI as shown in Fig. 1(b). For example, the effect would be considered if a spin current additionally breaks a directional symmetry of the system with broken inversion symmetry due to the structural reason. The effective field due to asymmetric DMI is

$$
\begin{aligned}
\vec{h}_{i,DM} &= \frac{1}{\mu_0 m}\left(\vec{S}_{i+x}\times\left(\vec{D}+\frac{\delta\vec{D}}{2}\right)-\vec{S}_{i-x}\times\left(\vec{D}-\frac{\delta\vec{D}}{2}\right)\right)\\
&\cong \frac{1}{\mu_0 m}\left(\left(\vec{S}_i+\frac{\partial \vec{S}_i}{\partial x}a\right)\times\left(\vec{D}+\frac{\delta\vec{D}}{2}\right)-\left(\vec{S}_i-\frac{\partial \vec{S}_i}{\partial x}a\right)\times\left(\vec{D}-\frac{\delta\vec{D}}{2}\right)\right) \qquad (2)\\
&= \vec{h}_{i,\vec{D}}+\frac{1}{\mu_0 m}\vec{S}_i\times\delta\vec{D},
\end{aligned}
$$

where $\delta\vec{D}$ is the additional vector caused by the asymmetry of the system. The additional effective field from $\delta\vec{D}$ is, therefore, $\delta\vec{h}_{i,\delta\vec{D}}=\frac{1}{\mu_0 m}\vec{S}_i\times\delta\vec{D}$. The $\delta\vec{h}_{i,\delta\vec{D}}$ is perpendicular to $\vec{S}_i$ as $\delta\vec{h}_{i,\delta J}$ is, and it makes the spin to be dynamic.

This additional effective field terms will attribute two additional torques in the Landau-Lifshitz (LL) equation. One term is related to a precession-like effect (spin rotation perpendicular to both the effective field and the spin direction) and another term is related to a damping effect (spin rotation within the plane of the effective field and spin direction). For general applications, we introduce another dimensionless parameter $r_{\Delta}$, allowing for the difference between the effects on precession and the damping motion generated by $\delta\vec{h}_{i,\Delta}$, where $\Delta = \delta J$ or $\delta\vec{D}$. The LL equation, including the additional effective fields, becomes

$$\frac{\partial \vec{m}_i}{\partial t} = -\gamma_L \vec{m}_i \times \vec{h}_{i,\delta=0} - \lambda \vec{m}_i \times \vec{m}_i \times \vec{h}_{i,\delta=0} - \sum_{\Delta} \left( r_{\Delta} \gamma_L \vec{m}_i \times \delta\vec{h}_{i,\Delta} + \lambda \vec{m}_i \times \vec{m}_i \times \delta\vec{h}_{i,\Delta} \right), \tag{3}$$

where $\vec{m}_i = m\vec{S}_i$, $\vec{h}_{i,\delta=0} = \vec{h}_{i,J} + \vec{h}_{i,\vec{D}}$, $\gamma_L = \mu_0 \gamma$, and $\lambda = \frac{\gamma_L \alpha}{m}$. The $\gamma$ and $\alpha$ are the electron gyromagnetic ratio and the Gilbert damping constant respectively. The third and fourth terms of Eq. (3) are reduced to

$$\begin{aligned}
r_{\delta J} \gamma_L \vec{m}_i \times \delta\vec{h}_{i,\delta J} &= r_{\delta J} \gamma_L \frac{a\,\delta J}{\mu_0 m^2} \left( \vec{m}_i \times \frac{\partial \vec{m}_i}{\partial x} \right) \\
\lambda \vec{m}_i \times \vec{m}_i \times \delta\vec{h}_{i,\delta J} &= -\lambda \frac{a\,\delta J}{\mu_0} \left( \frac{\partial \vec{m}_i}{\partial x} \right) \\
r_{\delta\vec{D}} \gamma_L \vec{m}_i \times \delta\vec{h}_{i,\delta\vec{D}} &= r_{\delta\vec{D}} \gamma_L \frac{1}{\mu_0 m^2} \left( \vec{m}_i \times \vec{m}_i \times \delta\vec{D} \right) \\
\lambda \vec{m}_i \times \vec{m}_i \times \delta\vec{h}_{i,\delta\vec{D}} &= -\frac{\lambda}{\mu_0} \left( \vec{m}_i \times \delta\vec{D} \right).
\end{aligned} \tag{4}$$

In Eq. (4), we used the properties that $\delta\vec{h}_{i,\delta J}$ is perpendicular to $\vec{S}_i$, because $\vec{S}_i$ varies while keeping its magnitude and $\vec{S}_i \perp \partial\vec{S}_i$.

Hence, the LL equation, including all the effective fields, becomes

$$\frac{\partial \vec{m}_i}{\partial t} = -\gamma_L \vec{m}_i \times \vec{h}_{i,\delta=0} - \lambda \vec{m}_i \times \vec{m}_i \times \vec{h}_{i,\delta=0} - r_{\delta J}\gamma_L \frac{a\,\delta J}{\mu_0 m^2}\left(\vec{m}_i \times \frac{\partial \vec{m}_i}{\partial x}\right) + \lambda \frac{a\,\delta J}{\mu_0}\left(\frac{\partial \vec{m}_i}{\partial x}\right)$$
$$- r_{\delta \vec{D}}\gamma_L \frac{1}{\mu_0 m^2}\left(\vec{m}_i \times \vec{m}_i \times \delta\vec{D}\right) + \frac{\lambda}{\mu_0}\left(\vec{m}_i \times \delta\vec{D}\right). \tag{5}$$

The Eq. (5) can be transformed into a form of the Landau-Lifshitz-Gilbert (LLG) equation as shown in Eq. (6),

$$\frac{\partial \vec{m}_i}{\partial t} = -\gamma_G \vec{m}_i \times \vec{h}_{i,\delta=0} + \frac{\alpha}{m}\vec{m}_i \times \frac{\partial \vec{m}_i}{\partial t} - u_{\delta J}\left(\frac{\partial \vec{m}_i}{\partial x}\right) + \beta_{\delta J} u_{\delta J}\left(\vec{m}_i \times \frac{\partial \vec{m}_i}{\partial x}\right) - \gamma_G \vec{m}_i \times \vec{h}_{FL}$$
$$+ \gamma_G \vec{m}_i \times \vec{m}_i \times \left(\beta_{\delta \vec{D}} \vec{h}_{FL}\right). \tag{6}$$

The relations among the parameters in Eq. (5) and Eq. (6) are

$$u_{\delta J} = \gamma_G \frac{a\,\delta J}{\mu_0 m}\frac{\alpha}{1+\alpha^2}\left(r_{\delta J} - 1\right), \tag{6.a}$$

$$\gamma_G \vec{h}_{FL} = \gamma_G \frac{\delta\vec{D}}{\mu_0 m}\frac{\alpha}{1+\alpha^2}\left(r_{\delta\vec{D}} - 1\right), \tag{6.b}$$

$$\beta_\Delta = \frac{1}{m\alpha}\left(\frac{r_\Delta + \alpha^2}{1 - r_\Delta}\right), \tag{6.c}$$

where $\Delta = \delta J$ or $\delta\vec{D}$, $\gamma_G = \gamma_L(1+\alpha^2)$.

In Eq. (6), the first two terms are the precession and damping terms in the original LLG equation. The third term in Eq. (6), $-u_{\delta J}\left(\frac{\partial \vec{m}_i}{\partial x}\right)$, corresponds to the adiabatic STT, where $u_{\delta J}$ is the adiabatic STT coefficient derived from our parameters in Eq. (6.a). It is the amplitude of the velocity vector $\vec{u}$ used in usual STT studies, and it is related to spin-polarized currents and given by $\vec{u} = \frac{\jmath P g \mu_B}{2em}\hat{\jmath}$, where $\jmath$ is the current density, $\hat{\jmath}$ is the direction of current flow, and $P$ its spin polarization rate. The parameters $\delta J$ and $r_{\delta J}$ have the relation $\alpha\left(r_{\delta J} - 1\right)\delta J = \frac{\hbar \jmath P}{2ae}$ with $\jmath$ and $P$. The fourth term in Eq. (6), $\beta_{\delta J} u_{\delta J}\left(\vec{m}_i \times \frac{\partial \vec{m}_i}{\partial x}\right)$, corresponds to the non-adiabatic STT. The dimensionless parameter $\beta_{\delta J}$ corresponds to the non-adiabatic spin torque component $\beta$ in usual STT study[2]. The origin and characterizing parameters of $\beta$ have been studied intensively since it has an important role in transverse and vortex domain wall dynamics[2, 16 - 18]. It is known that the $\beta$ value is affected by various factors such as momentum transfer[16,17], spin-flip scattering[18], and the magnetization gradient[14]. The experimentally measured $\beta$ has a wide range from 0 to $18\alpha$, with the possibility of being negative[19-24].

Our model also includes the cases where the coefficient of adiabatic torque term, $u_{\delta J}$, is zero and the coefficient of non-adiabatic torque term, $\beta_{\delta J} u_{\delta J}$, is non-zero under the condition $r_{\delta J} = 1$ and $\delta J \neq 0$. In case of DW motion studies, this condition means that the pure non-adiabatic behavior of DW motion will show when the directional preference of exchange interaction occurs for any reason without the spin polarized current. We guess that this phenomenon can observed in the two-dimensional magnetic insulator system, such as TmIG film system, with external electric field applied in planar direction, which can generate electrical polarization in the system.

The last two terms of Eq. (6), $-\gamma_G \vec{m}_i \times \vec{h}_{FL}$ and $\gamma_G \vec{m}_i \times \vec{m}_i \times \left(\beta_{\delta \vec{D}} \vec{h}_{FL}\right)$, are called field-like and anti-damping torque terms in usual SOT study. These are related to the torques induced by the intrinsic or extrinsic spin-orbit torque[10]. In our approach, if no specific situation is considered, $\delta \vec{D}$ can have any three-dimensional direction, so these terms can produce spin dynamics behavior such as the Rashba effect or spin Hall effect, depending on the proper choice of $\delta \vec{D}$ direction. For example, the Rashba Hamiltonian resulting inversion symmetry breaking in the direction perpendicular to the two-dimensional plane is known as $\mathcal{H}_R = \alpha_R \left(\vec{\sigma} \times \vec{k}\right) \cdot \hat{z}$, where $\alpha_R$ is the Rashba coupling constant, $\hbar \vec{k}$ is the electron's momentum and $\vec{\sigma}$ is the Pauli matrix vector[5]. The effective field of Rashba effect becomes $\vec{h}_R = \frac{\alpha_R}{\mu_0 m} \left(\vec{k} \times \hat{z}\right)$, and it is known that the field generates field-like torques dominantly. In this case, we can understand the Rashba effect using the $r_{\delta \vec{D}} = -\alpha^2$ and $-\alpha\, \delta \vec{D} = \alpha_R \left(\vec{k} \times \hat{z}\right)$ conditions, which make $\vec{h}_{FL}$ become the effective field of Rashba effect. As same way, the case that anti-damping torque term is mainly considered, such as spin hall effect case, also can be interpreted using $r_{\delta \vec{D}} = 1$ with proper $\delta\vec{D}$ choice. If the real system should consider both field-like and anti-damping terms, choosing a suitable value of $r_{\delta\vec{D}}$ provides the ratio between two torque terms.

In addition, there is no reason that $\delta J$ and $\delta\vec{D}$ are restricted to be constants; they may be functions of external conditions breaking inversion symmetry in the system. For example, our model can be expanded to use for the magneto-elastic effect, which is generated by lattice vibrations through $\delta J$, with $\delta\vec{D}$ having the form of $\sim e^{i\omega t}$. In a similar way, the approach using our model can be extended to more general causes of breaking inversion symmetry in systems such as an external electric field, surface polarization, a surficial or interfacial spin environment, spontaneous symmetry breaking, etc.

In our approach, we obtained those additional terms, not from the mechanism of physical origins, but from the symmetry discussion. We do not confine our model to a special case, like STT or SOT, so we could find the unified form of additional torque term induced by inversion symmetry breaking as shown in Eq. (7).

$$\vec{\tau} = -\gamma_G \left(\frac{\alpha}{1+\alpha^2}\right) \sum_{\Delta} (r_\Delta - 1)\left(\delta\vec{h}_{i,\Delta} - \beta_\Delta \vec{m}_i \times \delta\vec{h}_{i,\Delta}\right). \tag{7}$$

This unified form can be applied on various subjects by choosing a proper $\delta\vec{h}_{i,\Delta}$.

## III. SUMMARY

We discussed the effect of broken inversion symmetry on spin dynamics. We extended the LLG equation with asymmetric and directional spin-spin interactions, including both the exchange interaction and the DMI. The asymmetric and directional interaction results in additional torque terms related to the STT and SOT. In developing this theory, we do not require an understanding of the physical origins of the parameters; we only consider the effects of the broken inversion symmetry. Therefore, STT and SOT are the natural consequences of the broken inversion symmetry by a spin current or electric field. Our model offers an insightful understanding of the spin dynamics generated by broken inversion symmetry, and opens up a new perspective for us that STT and SOT can be interpreted as the directional preference of spin-spin interaction.

## ACKNOWLEDGMENTS

This research was supported by a Grant from the National Research Foundation of Korea, funded by the Korean Government (2015R1D1A1A01056971).

---

## REFERENCES

[1] N. Vernier, D. A. Allwood, D. Atkinson, M. D. Cooke, and R. P. Cowburn, *Europhys. Lett.* **65**, 526 (2004).

[2] A. Thiaville, Y. Nakatani, J. Miltat, and Y. Suzuki, Europhys. Lett. **69** (6), 990 (2005)

[3] O. A. Tretiakov and Ar. Abanov, Phys. Rev. Lett. **105**, 157201 (2010).

[4] I. M. Miron, T. Moore, H. Szambolics, L. D. Buda-Prejbeanu, S. Auffret, B. Rodmacq, S. Pizzini, J. Vogel, M. Bonfim, A. Schuhl and G. Gaudin, Nat. Mat., **10**, 419 (2011)

[5] E. Martinez, S. Emori, and S. D. Beach, Appl. Phys. Lett. **103**, 072406 (2013)

[6] V. P. Kravchuk, J. Magn. Magn. Mater. **367**, 9 (2014).

[7] L. Liu, O. J. Lee, T. J. Gudmundsen, D. C. Ralph, and R. A. Buhrman, Phys. Rev. Lett. **109**, 096602 (2012).

[8] A. Manchon and S. Zhang, *Phys. Rev. B* **79**, 094422 (2009)

[9] L. Liu, C. F. Pai, Y. Li, H. W. Tseng, D. C. Ralph, and R. A. Buhrman, Science **336**(6081), 555 (2012)

[10] A. V. Khvalkovskiy, V. Cros, D. Apalkov, V. Nikitin, M. Krounbi, K. A. Zvezdin, A. Anane, J. Grollier, and A. Fert, Phys. Rev. B **87**, 020402(R) (2013)

[11] A. V. Khvalkovskiy, K. A. Zvezdin, Ya. V. Gorbunov, V. Cros, J. Grollier, A. Fert, and A. K. Zvezdin Phys. Rev. Lett. **102**, 067206 (2009)

[12] A. Manchon, arXiv:1204.4869 (2012)

[13] I. Dzyaloshinsky, J. Phys. Chem. Solids **4**, 241 (1958).

[14] T. Moriya, Phys. Rev. **120**, 91 (1960).

[15] X. X. Zhang and N. Nagaosa, arXiv:cond-mat/1608.06362.(2016)

[16] C. Burrowes, A. P. Mihai, D. Ravelosona, J.-V. Kim, C. Chappert, L. Vila, A. Marty, Y. Samson, F. Garcia-Sanchez, L. D. Buda-Prejbeanu, I. Tudosa, E. E. Fullerton and J.-P. Attané, Nat. Phys. **6**, 17 (2010).

[17] M. Eltschka, M. Wotzel, J. Rhensius, S. Krzyk, U. Nowak, and M. Klaui, Phys. Rev. Lett. **105**, 056601 (2010).

[18] S.D. Pollard, L. Huang, K.S. Buchanan, D.A. Arena and Y. Zhu, Nat. Comm. **3**, 1028 (2012).

[19] G. Tatara and H. Kohno, Phys. Rev. Lett. **92**, 086601 (2004).

[20] G. Tatara, H. Kohno, and J. Shibata, J. Phys. Soc. Jpn. **77**, 031003 (2008).

[21] S. Zhang and Z. Li, Phys. Rev. Lett. **93**, 127204 (2004).

[22] I. Garate, K. Gilmore, M. D. Stiles, and A. H. MacDonald, Phys. Rev. B **79**, 104416 (2009)

[23] S. Bohlens and D. Pfannkushe, Phys. Rev. Lett. **105**, 177201 (2010)

[24] S. G. Je, S. C. Yoo, J. S. Kim, Y. K. Park, M. H. Park, J. Moon, B. C. Min, and S. B. Choe, Phys. Rev. Lett. **118**, 167205 (2017)

FIGURE CAPTIONS

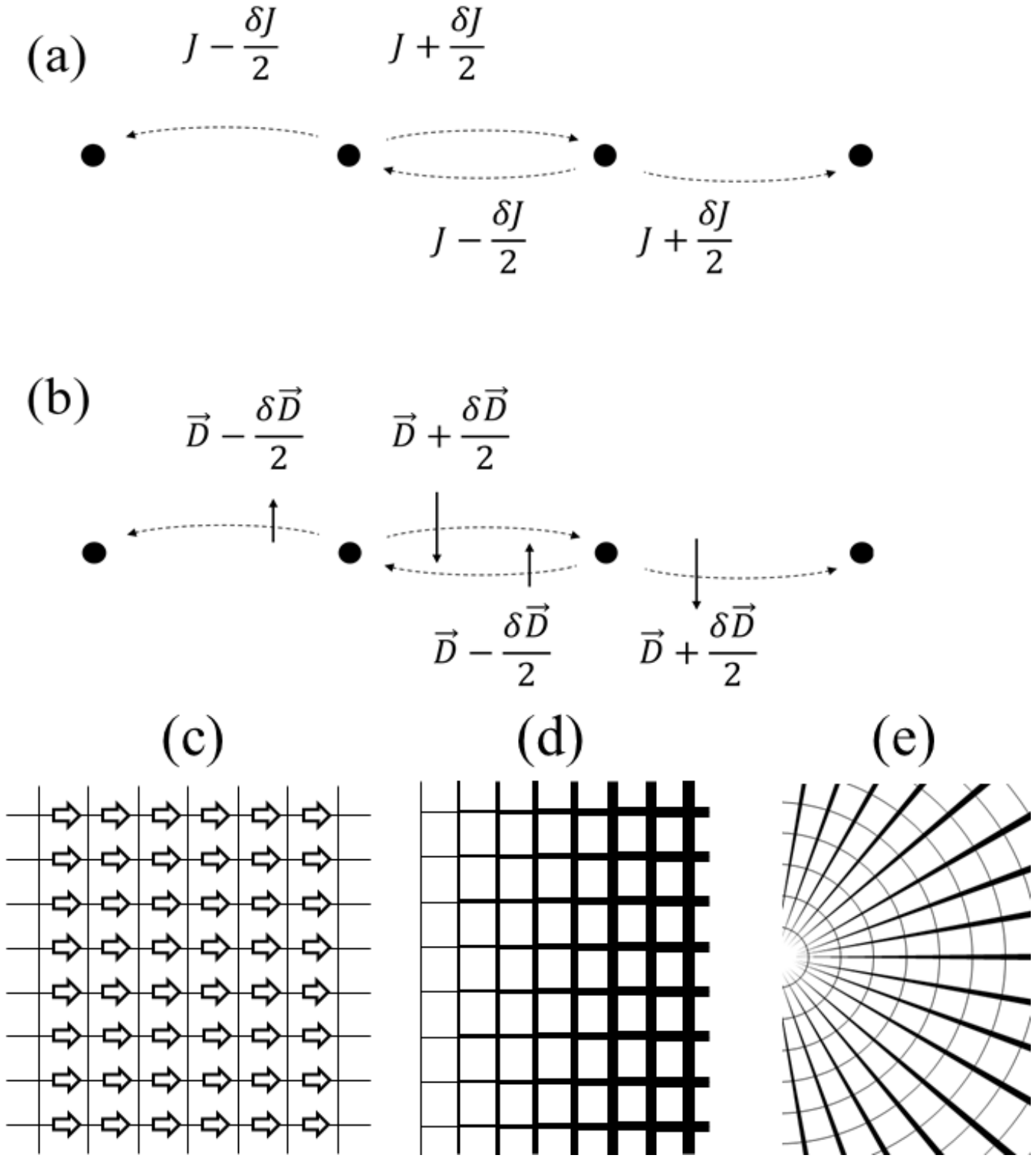

FIG. 1. Schematic illustration of the asymmetric interaction between nearest neighbor spins for (a) exchange interaction and (b) DMI. (c)-(e) shows the examples in which the asymmetric and directional interaction may occur in grid models. (c) The existence of fields or current, (b) edge or boundary of the system, or (e) non-symmetric structure or distortion can be the reasons of asymmetric or directional spin-spin interaction as in (a) and (b).